\documentclass[sn-basic,Numbered]{sn-jnl}

\usepackage{graphicx}%
\usepackage{amsmath,amssymb,amsfonts}%
\usepackage{amsthm}%
\usepackage{mathrsfs}%
\usepackage[title]{appendix}%
\usepackage{xcolor}%
\usepackage{textcomp}%
\usepackage{manyfoot}%
\usepackage{booktabs}%
\usepackage{algorithm}%
\usepackage{algorithmicx}%
\usepackage{algpseudocode}%
\usepackage{listings}%

\usepackage{colortbl}
\usepackage{multicol}
\usepackage{multirow}
\usepackage{siunitx}
\usepackage{caption,subcaption}
\usepackage{tabularx}
\usepackage{etoolbox}
\usepackage{pgf} 

\definecolor{low}{HTML}{b77495}  
\definecolor{high}{HTML}{3a0e37}  
\definecolor{white}{HTML}{ffffff}
\newcommand*{\opacity}{85}
\newcommand*{\minvala}{0.7}
\newcommand*{\maxvala}{0.78}
\newcommand*{\minvalb}{0.59}
\newcommand*{\maxvalb}{0.82}
\newcommand*{\minvalca}{0.45}
\newcommand*{\maxvalca}{0.86}
\newcommand*{\minvalcb}{0.60}
\newcommand*{\maxvalcb}{0.81}
\newcommand*{\minvalcc}{0.71}
\newcommand*{\maxvalcc}{0.92}
\newcommand*{\minvalcd}{0.52}
\newcommand*{\maxvalcd}{0.80}
\newcommand*{\minvalce}{0.62}
\newcommand*{\maxvalce}{0.87}
\newcommand*{\minvalda}{0.81}
\newcommand*{\maxvalda}{0.97}
\newcommand*{\minvaldb}{0.53}
\newcommand*{\maxvaldb}{0.78}
\newcommand*{\minvaldc}{0.70}
\newcommand*{\maxvaldc}{0.94}
\newcommand*{\minvaldd}{0.62}
\newcommand*{\maxvaldd}{0.79}
\newcommand*{\minvalde}{0.55}
\newcommand*{\maxvalde}{0.78}
\newcommand{\grada}[1]{
    \ifdimcomp{#1pt}{>}{\maxvala pt}{#1}{
        \ifdimcomp{#1pt}{<}{\minvala pt}{#1}{
            \pgfmathparse{int(round(100*(#1/(\maxvala-\minvala))-(\minvala*(100/(\maxvala-\minvala)))))}
            \xdef\tempa{\pgfmathresult}
            \cellcolor{high!\tempa!low!\opacity}#1}}
}
\newcommand{\gradb}[1]{
    \ifdimcomp{#1pt}{>}{\maxvalb pt}{#1}{
        \ifdimcomp{#1pt}{<}{\minvalb pt}{#1}{
            \pgfmathparse{int(round(100*(#1/(\maxvalb-\minvalb))-(\minvalb*(100/(\maxvalb-\minvalb)))))}
            \xdef\tempa{\pgfmathresult}\cellcolor{high!\tempa!low!\opacity}#1}}
}
\newcommand{\gradca}[1]{
    \ifdimcomp{#1pt}{>}{\maxvalca pt}{#1}{
        \ifdimcomp{#1pt}{<}{\minvalca pt}{#1}{
            \pgfmathparse{int(round(100*(#1/(\maxvalca-\minvalca))-(\minvalca*(100/(\maxvalca-\minvalca)))))}
            \xdef\tempa{\pgfmathresult}
            \cellcolor{high!\tempa!low!\opacity}#1}}
}
\newcommand{\gradcb}[1]{
    \ifdimcomp{#1pt}{>}{\maxvalcb pt}{#1}{
        \ifdimcomp{#1pt}{<}{\minvalcb pt}{#1}{
            \pgfmathparse{int(round(100*(#1/(\maxvalcb-\minvalcb))-(\minvalcb*(100/(\maxvalcb-\minvalcb)))))}
            \xdef\tempa{\pgfmathresult}
            \cellcolor{high!\tempa!low!\opacity}#1}}
}
\newcommand{\gradcc}[1]{
    \ifdimcomp{#1pt}{>}{\maxvalcc pt}{#1}{
        \ifdimcomp{#1pt}{<}{\minvalcc pt}{#1}{
            \pgfmathparse{int(round(100*(#1/(\maxvalcc-\minvalcc))-(\minvalcc*(100/(\maxvalcc-\minvalcc)))))}
            \xdef\tempa{\pgfmathresult}
            \cellcolor{high!\tempa!low!\opacity}#1}}
}
\newcommand{\gradcd}[1]{
    \ifdimcomp{#1pt}{>}{\maxvalcd pt}{#1}{
        \ifdimcomp{#1pt}{<}{\minvalcd pt}{#1}{
            \pgfmathparse{int(round(100*(#1/(\maxvalcd-\minvalcd))-(\minvalcd*(100/(\maxvalcd-\minvalcd)))))}
            \xdef\tempa{\pgfmathresult}
            \cellcolor{high!\tempa!low!\opacity}#1}}
}
\newcommand{\gradce}[1]{
    \ifdimcomp{#1pt}{>}{\maxvalce pt}{#1}{
        \ifdimcomp{#1pt}{<}{\minvalce pt}{#1}{
            \pgfmathparse{int(round(100*(#1/(\maxvalce-\minvalce))-(\minvalce*(100/(\maxvalce-\minvalce)))))}
            \xdef\tempa{\pgfmathresult}
            \cellcolor{high!\tempa!low!\opacity}#1}}
}
\newcommand{\gradda}[1]{
    \ifdimcomp{#1pt}{>}{\maxvalda pt}{#1}{
        \ifdimcomp{#1pt}{<}{\minvalda pt}{#1}{
            \pgfmathparse{int(round(100*(#1/(\maxvalda-\minvalda))-(\minvalda*(100/(\maxvalda-\minvalda)))))}
            \xdef\tempa{\pgfmathresult}
            \cellcolor{high!\tempa!low!\opacity}#1}}
}
\newcommand{\graddb}[1]{
    \ifdimcomp{#1pt}{>}{\maxvaldb pt}{#1}{
        \ifdimcomp{#1pt}{<}{\minvaldb pt}{#1}{
            \pgfmathparse{int(round(100*(#1/(\maxvaldb-\minvaldb))-(\minvaldb*(100/(\maxvaldb-\minvaldb)))))}
            \xdef\tempa{\pgfmathresult}
            \cellcolor{high!\tempa!low!\opacity}#1}}
}
\newcommand{\graddc}[1]{
    \ifdimcomp{#1pt}{>}{\maxvaldc pt}{#1}{
        \ifdimcomp{#1pt}{<}{\minvaldc pt}{#1}{
            \pgfmathparse{int(round(100*(#1/(\maxvaldc-\minvaldc))-(\minvaldc*(100/(\maxvaldc-\minvaldc)))))}
            \xdef\tempa{\pgfmathresult}
            \cellcolor{high!\tempa!low!\opacity}#1}}
}
\newcommand{\graddd}[1]{
    \ifdimcomp{#1pt}{>}{\maxvaldd pt}{#1}{
        \ifdimcomp{#1pt}{<}{\minvaldd pt}{#1}{
            \pgfmathparse{int(round(100*(#1/(\maxvaldd-\minvaldd))-(\minvaldd*(100/(\maxvaldd-\minvaldd)))))}
            \xdef\tempa{\pgfmathresult}
            \cellcolor{high!\tempa!low!\opacity}#1}}
}
\newcommand{\gradde}[1]{
    \ifdimcomp{#1pt}{>}{\maxvalde pt}{#1}{
        \ifdimcomp{#1pt}{<}{\minvalde pt}{#1}{
            \pgfmathparse{int(round(100*(#1/(\maxvalde-\minvalde))-(\minvalde*(100/(\maxvalde-\minvalde)))))}
            \xdef\tempa{\pgfmathresult}
            \cellcolor{high!\tempa!low!\opacity}#1}}
}

\newcommand{\reffig}[1]{Fig.\,\ref{#1}}

\makeatletter
\def\BState{\State\hskip-\ALG@thistlm}
\makeatother

\begin{document}

\title[Article Title]{Voxel-wise classification for porosity investigation of additive manufactured parts with 3D unsupervised and (deeply) supervised neural networks}


\author*[1,6]{\fnm{Domenico} \sur{Iuso}}\email{diuso@pm.me}
\author[2,3]{\fnm{Soumick} \sur{Chatterjee}}
\author[4]{\fnm{Sven} \sur{Cornelissen}}
\author[5]{\fnm{Dries} \sur{Verhees}}
\author[1,6]{\fnm{Jan} \sur{De Beenhouwer}}
\equalcont{These authors contributed equally to this work.}
\author[1,6]{\fnm{Jan} \sur{Sijbers}}
\equalcont{These authors contributed equally to this work.}

\affil*[1]{\orgdiv{imec-Vision Lab, Department of Physics}, \orgname{University of Antwerp}, \orgaddress{\city{Antwerp}, \postcode{2610}, \country{Belgium}}}
\affil[2]{\orgdiv{Faculty of Computer Science}, \orgname{Otto von Guericke University}, \orgaddress{\street{Street}, \city{Magdeburg}, \postcode{39106}, \country{Germany}}}
\affil[3]{\orgdiv{Genomics Research Centre}, \orgname{Human Technopole}, \orgaddress{\street{V.le Rita Levi-Montalcini 1}, \city{Milan}, \postcode{20157}, \country{Italy}}}
\affil[4]{\orgname{Materialise NV.}, \orgaddress{\street{Technologielaan 15}, \city{Leuven}, \postcode{3001}, \country{Belgium}}}
\affil[5]{\orgname{Flanders Make vzw.}, \orgaddress{\street{Oude Diestersebaan 133}, \city{Lommel}, \postcode{3920}, \country{Belgium}}}
\affil[6]{\orgname{DynXlab: Center for 4D Quantitative X-ray Imaging and Analysis}, \orgaddress{\city{Antwerp}, \postcode{2610}, \country{Belgium}}}


\abstract{Additive Manufacturing (AM) has emerged as a manufacturing process that allows the direct production of samples from digital models. To ensure that quality standards are met in all manufactured samples of a batch, X-ray computed tomography (X-CT) is often used combined with automated anomaly detection. For the latter, deep learning (DL) anomaly detection techniques are increasingly, as they can be trained to be robust to the material being analysed and resilient towards poor image quality. Unfortunately, most recent and popular DL models have been developed for 2D image processing, thereby disregarding valuable volumetric information.

This study revisits recent supervised (UNet, UNet++, UNet 3+, MSS-UNet) and unsupervised (VAE, ceVAE, gmVAE, vqVAE) DL models for porosity analysis of AM samples from X-CT images and extends them to accept 3D input data with a 3D-patch pipeline for lower computational requirements, improved efficiency and generalisability. The supervised models were trained using the Focal Tversky loss to address class imbalance that arises from the low porosity in the training datasets. The output of the unsupervised models is post-processed to reduce misclassifications caused by their inability to adequately represent the object surface. The findings were cross-validated in a 5-fold fashion and include: a performance benchmark of the DL models, an evaluation of the post-processing algorithm, an evaluation of the effect of training supervised models with the output of unsupervised models. In a final performance benchmark on a test set with poor image quality, the best performing supervised model was UNet++ with an average precision of 0.751 $\pm$ 0.030, while the best unsupervised model was the post-processed ceVAE with 0.830 $\pm$ 0.003. The VAE/ceVAE models demonstrated superior capabilities, particularly when leveraging post-processing techniques.}




\maketitle

%
%
\section{Introduction}
Additive manufacturing is gaining interest since it is a low-waste production technique that can conveniently produce complex objects from a given CAD file~\cite{frazier2014metal}. The latest developments in 3D printing technology allow to print metallic alloys effectively, as in the case of Selective Laser Melting~\cite{jia2021scanning}. While this technique has many advantages, a key challenge is printing metallic alloys without defects. The mechanical behaviour of 3D printed parts, including tensile or fatigue stress behaviours, greatly depends on their overall structural integrity~\cite{tang2017oxides}. Defects such as common keyhole or lack-of-fusion pores~\cite{zhang2017defect} can seriously degrade the mechanical properties of printed parts by becoming initiation centres for crack development. 

For non-destructive evaluation of the printing process and quality assurance of the printed part, X-CT is often employed~\cite{sarkon2022state}. X-CT has been used to analyse the internal and external surface properties~\cite{leary2016selective, salarian2018use}, the structural integrity of samples~\cite{Thompson_2016}, as well as identification and quantification of the number of defects that arise from the AM process~\cite{SANAEI2021100724, vandecasteele2023towards}.

Detecting anomalies from X-CT data is a challenging task (due to, for example, inhomogeneous density of the sample, a low contrast-to-noise ratio, or beam hardening artefacts) that may lead to incorrect classification. For such a task, data-driven deep learning-based approaches have been shown to outperform traditional machine learning techniques because they can better handle complex and varied definitions of anomalies~\cite{bihani2022mudrocknet, kim2022achieving, wong2021automatic}. Anomalies can be detected in a supervised or unsupervised fashion. While supervised methods require an annotated data set, unsupervised methods are more desirable because the training data need not be annotated. Apart from reducing the technical overhead for the user due to the gathering of annotated data, unsupervised models nullify the impact that noisy annotations have on the performance of the model. On the other hand, the general challenge that researchers of unsupervised approaches face are the high recall rate and/or low precision when these approaches are compared to their supervised counterpart~\cite{yang2021visual}. 

The majority of studies on voxel-wise classification tasks with DL techniques are focused on the analysis of a stack of 2D images~\cite{ar2020segmentation, wang2022deep, fend2021reconstruction, bouget2019semantic}. For voxel-wise classification of pores in AM samples, a 2D approach suboptimal as small pores usually span only a few voxels in the three directions in X-CT images and suffer from a low contrast-to-noise ratio. Moreover, pores may be elongated as they usually exhibit anisotropy~\cite{maskery2016quantification}, with a high risk of being ignored by 2D pixel-wise classification methods. Hence, there is a need for 3D pore detection models, as recently suggested by Wong et al.~\cite{wong2021automatic}. Their initial study with a simple model architecture (UNet) showed promising results using a 3D approach, but deep supervision has not been explored. Deep supervision can yield more reliable results since the hidden layers of the models are enticed to comply with the desired output~\cite{li2022comprehensive}. However, training supervised models directly on a dataset with reduced porosity may seriously affect detection performance due to a strong class imbalance between the number of voxels that belong to pores and those that do not~\cite{bria2020addressing}. Moreover, supervised models are known to be highly sensitive to training labels. 
Unsupervised models, especially those based on VAE architectures, suffer a well-known issue of blurry representation of input images, because the models learn a low-dimensional representation that may not capture fine details~\cite{guo2020variational}. For these models, the difference between the input and output images alone, which is usually the voxel-wise anomaly score, may not be a good indication of anomaly presence. Therefore, the voxel-wise anomaly score can be enhanced with a more complex anomaly score or dedicated post-processing~\cite{zimmerer2018context, baur2021autoencoders}. 

In this work, 2D supervised and unsupervised DL models are first revisited and subsequently extended to 3D for voxel-wise classification of pores on X-CT samples of varying alloys. Through a 3D patch-based approach and data augmentation, the classification aims to be independent of the material and shape of the AM samples. Several deeply supervised models are trained (MSS-UNet~\cite{ZHAO2020100357}, UNet++~\cite{zhou2018unet++}, and UNet 3+~\cite{huang2020unet}) while a more traditional UNet~\cite{ronneberger2015u} serves as a baseline to compare the remaining three architectures, which are composed of similar building blocks to easier the comparison. To address the class imbalance that arises from the low amount of defects, the models are trained with the Focal Tversky (FTL) function, which allows models to penalise anomalies more effectively~\cite{abraham2019novel}. Since the FTL is a parametric function, the optimal parameters were found with a parameter search. The roster of unsupervised models (VAE~\cite{kingma2013auto}, ceVAE~\cite{zimmerer2018context}, gmVAE~\cite{dilokthanakul2016deep}, and vqVAE~\cite{van2017neural}) aims to compare older and novel complex model architectures. To reduce misclassifications, the anomaly score of these models is post-processed, due to the inability of these models to represent the object surface adequately. Finally, the supervised models are trained again with the post-processed output of an unsupervised model instead of (potentially) noisy annotations, to evaluate the impact on the performance of the supervised models.  On the best performing model of all the experiments, an experiment was run to assess the decrease in performance when lowering the number of X-ray projections and exposure. In summary, our study has three main contributions:
\begin{itemize}
\item Cross-validated evaluation of multiple supervised and unsupervised DL models is conducted using a 3D patch-based approach, enabling a comparable assessment of their performance.
\item A post-processing algorithm is proposed and evaluated to address the issue of blurry image representation in VAE models. 
\item The impact of using unsupervised model labels instead of heuristic algorithm labels for training the DL models is evaluated.
\end{itemize}

%
%
\section{Materials}
\label{sec:materials}
\begin{figure}
\centering
\begin{subfigure}{.28\textwidth}
  \includegraphics[width=\textwidth]{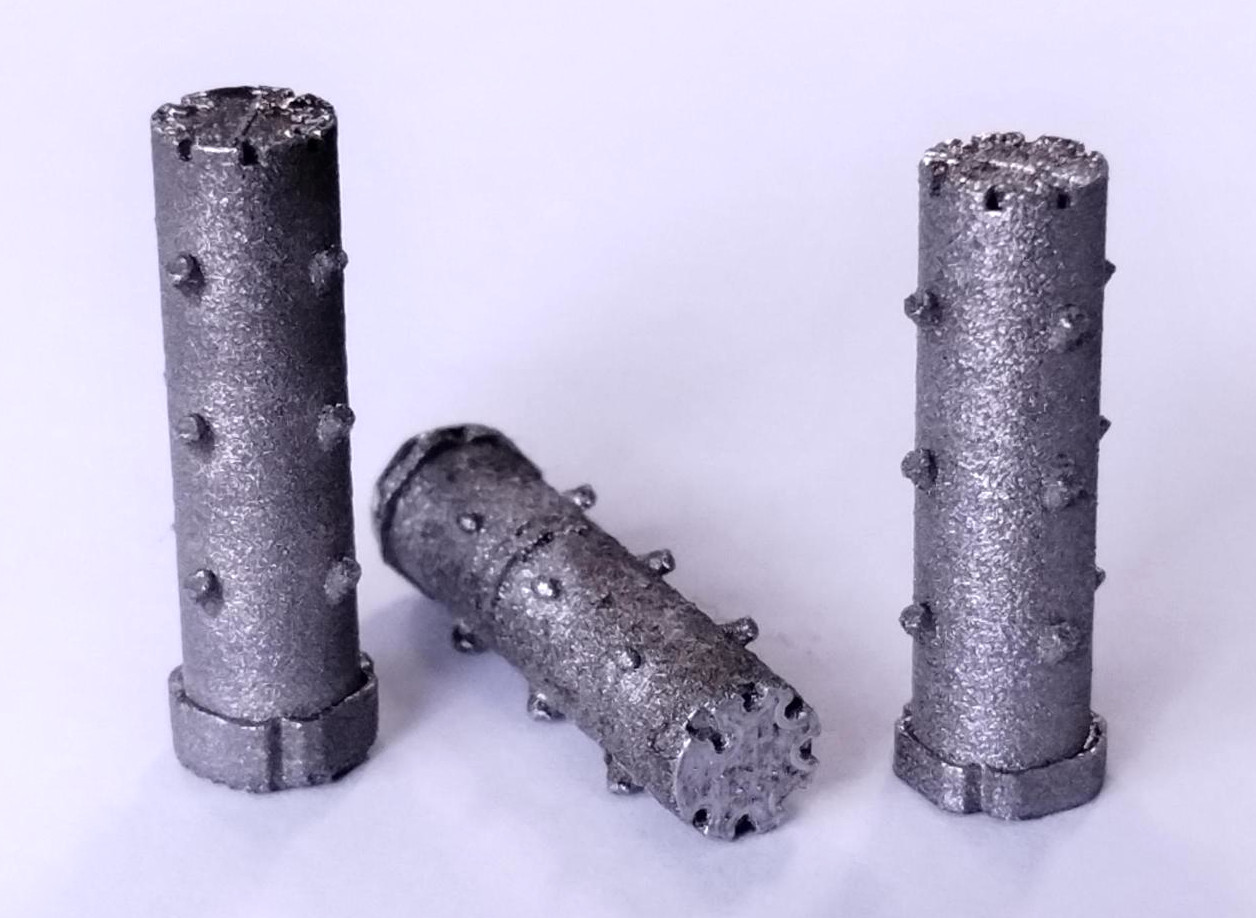}
\end{subfigure}%
\caption{Some samples used in this study. From left to right, a stainless steel 316L (CLM), a CoCr-DG1 (PLM), and a TiAl6V4 (CLM) sample.}
\label{fig:photo-samples}
\end{figure}

Various DL models for voxel-wise classification of pores were trained using 3D X-ray CT images of AM samples. To this end, AM samples were manufactured through the selective laser melting process, in a continuous (CLM,~\cite{meiners1998shaped}) or pulsed laser melting (PLM,~\cite{abe2001manufacturing}) strategy. Five cylindrical samples of different materials were 3D printed (as shown in Fig. \ref{fig:photo-samples}): one with TiAl6V4, two with CoCr-DG1 alloy and two with stainless steel 316L. Printing the test objects in multiple materials allowed to assess the effectiveness of voxel-wise pore classification across different materials. In the CAD model used for the 3D printing, the cylinders had an eight of 20 mm and a diameter of 5 mm. In addition to the cylindrical samples, a stainless steel 316L cube with an edge length of 9 mm was also printed. The cube was specifically printed to provide an object with different shape and poorer X-CT image quality, which is useful for evaluating porosity in a challenging visual environment and to ensure that DL models are not learning information regarding the shape of the object. These samples were essential for this study as their X-CTs provided the digital dataset with which the neural networks could be trained to classify the porosity. Porosity was intentionally induced in all samples using controlled laser parameters, as described in~\cite{booth2022encoding}.

\begin{figure}
\hfill
\begin{minipage}{.55\textwidth}
\centering
\includegraphics[width=0.96\textwidth, keepaspectratio]{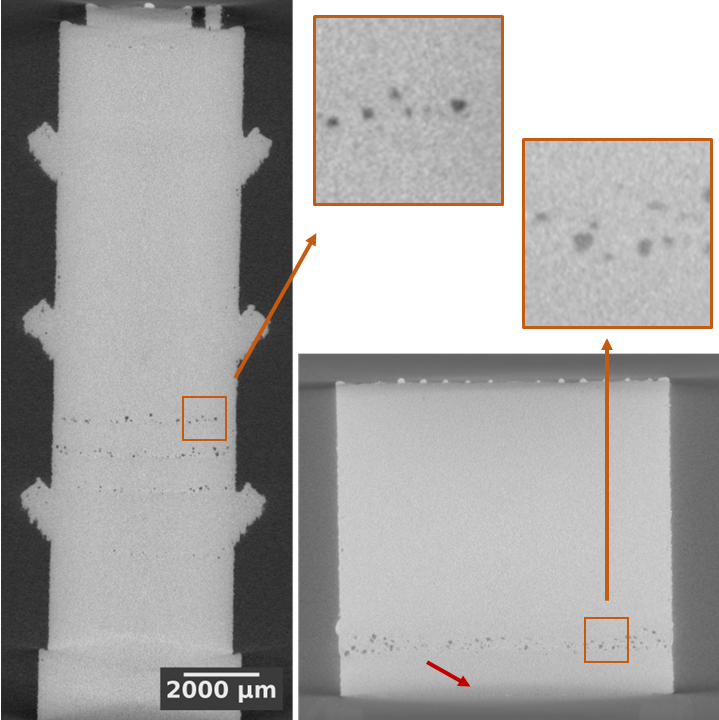}
\end{minipage}
\begin{minipage}{.42\textwidth}
\centering
\subcaptionbox{}{\includegraphics[width=0.72\textwidth]{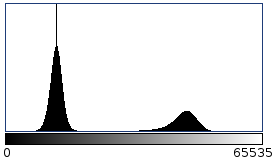}}
\\[1\baselineskip]
\subcaptionbox{}{\includegraphics[width=0.72\textwidth]{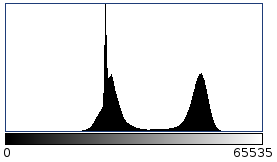}}
\end{minipage}\hfill
\caption{A slice of the X-CT of a cylindrical sample (left) and of the cubic sample (right), with equal colour-map and scale. While all the X-CT of cylindrical samples share similar image quality, the cube has stronger artefacts (which are particularly visible at the extremities of the cube) and consequently less contrast. The histograms (a) and (b) refer to the cylindrical sample and of the cubic volume, respectively. The two peaks in each histogram are related to the background (lower) and foreground (higher) colours. The quality of each sample is defined by its distance between the peaks and the broadness of the bells, which are influenced by artefacts and noise.}\label{fig:samples-and-histograms}
\end{figure}

Next, 3D images of the AM samples were generated by scanning them with a micro-CT X-ray system~\cite{Samber21} and reconstructed with the FDK algorithm~\cite{feldkamp1984practical} with a 10 $\mu$m resolution.  The imaging settings, such as filament power, peak kV of the anode, exposure time, source filter, etc., were selected for each cylindrical sample to ensure comparable image quality. However, the geometrical distances and the number of projections were kept constant for all cylindrical samples, with a source-to-detector distance (SDD) of 650.0 mm, a source-to-object distance (SOD) of 43.33 mm, and 4283 projections. The cubic sample was scanned with different SDD (950.0 mm) and SOD (63.33 mm) and had a lower number of X-ray projections (2878) than the other scans. The X-CT of the cubic sample was also affected by severe cone-beam artefacts and poor beam-hardening compensation. The cubic sample was particularly challenging due to its different geometry and visual environment (as noticeable in Fig. \ref{fig:samples-and-histograms}), making it useful for evaluating porosity.

%
%
\section{Methods}
Several DL models were trained for classification of porosity from X-CTs of the AM samples at voxel level. The X-CT images were organised into training, validation, and testing sets, as explained in section \ref{sec:methods:datasets}. The models were trained using either supervised or unsupervised approaches, which are further discussed in section \ref{sec:methods:networks}, and all models were trained using a common training framework, as described in section \ref{sec:methods:training}.

For the training of supervised models, the class imbalance of labels was addressed using the FTL function, which will be discussed in section \ref{sec:methods:FTL}. The class imbalance arose from the low amount of pores (positive instance of labels) within the training dataset. After training, and only for the unsupervised models, the anomaly score is post-processed, as unsupervised models are known to produce blurry representations of the input. The post-processing procedure is explained in section \ref{sec:methods:post-process}.

\subsection{Dataset}
\label{sec:methods:datasets}
The X-CT images of several AM samples composed the digital dataset for training, validation, and testing of the DL models. In a 5-fold manner, the X-CT images of the cylindrical samples were organised into 4 samples for the train-set and 1 sample for the validation-set. Noise, image artefacts, and misclassified voxel-wise labels (commonly referred to as 'noisy labels') can negatively affect training and lead to inaccurate predictions. To mitigate the influence of noisy labels during training and to expand the training sets, data augmentation was employed~\cite{song2022learning}. 
The data augmentation created novel spatial configurations by flipping of patches in random directions and elastic distortion while teaching the networks to be resilient against noise, specific attenuation of samples, and artefacts such as cone-beam and beam-hardening. After data augmentation was applied at every training epoch to each of the cylindrical samples, which have around 800x800x2000 voxels, 3D patches of 64x64x64 voxels were extracted and supplied to the neural networks.

To assign a label to each voxel of the X-CTs that indicates  whether it is a pore or not, a 3D postprocessing algorithm was applied. 
The high-level pseudo-code in Algorithm \ref{alg:pore-seg} outlines the pore identification process.
\begin{algorithm}
\caption{Pore-mask extraction}
\label{alg:pore-seg}
\begin{algorithmic}[1]
\State $input := CT volume$
\Procedure{Object masking}{input}
\State $input_{thr} \gets input > Otsu(input)$
\State $obj\_mask \gets FloodFill(input_{thr})$
\EndProcedure
\Procedure{Pore masking}{input, obj\_mask}
\State $pore_{mask} \gets input < Otsu(input \in obj\_mask)$
\State $pore\_list \gets ConnectivityFilter(pore_{mask})$
\EndProcedure
\Procedure{Noise removal}{pore\_list}
\ForAll {pore in pore\_list}
\If {$Dim(pore) < min\_dims$} 
\State \textbf{remove} pore
\EndIf
\EndFor
\State $pore_{mask} \gets ConvertToVolume(pore\_list)$
\EndProcedure
\State $output := pore_{mask}$
\end{algorithmic}
\end{algorithm}
First, a watertight mask of the sample is created (1\textsuperscript{st} procedure in Alg.~\ref{alg:pore-seg}) to exclude values outside the sample, which could bias the Otsu threshold calculation due to the CT scanner's larger field of view. A FloodFill algorithm was subsequently applied to obtain the outline of the object, after which a second Otsu thresholding separated the foreground (pores) from the background (solid material). To reduce the risk of voxel misclassification due to imaging noise, pore-voxels in a 6-connected 3D neighbourhood were rejected if their boundary box was smaller than 2 in at least one dimension 
\cite{DUPLESSIS20181102, kim2017investigation}. Any residual misclassification due to partial-volume effects and imaging artefacts contributed to the noise of the labels. 

Accurately and reliably labelling the X-CT scan of the cubic sample was a challenging task due to its poor image quality, as discussed in section \ref{sec:materials}. Given the limitations of automated voxel-wise annotation, manual labelling was the only viable option to achieve the desired level of accuracy and dependability in the labels.

\subsection{Deep learning models}
\label{sec:methods:networks}
The study used two types of models: VAE-based models (VAE~\cite{kingma2013auto}, ceVAE~\cite{zimmerer2018context}, gmVAE~\cite{dilokthanakul2016deep}, and vqVAE~\cite{van2017neural}) and UNet-based models (UNet~\cite{ronneberger2015u}, MSS-UNet~\cite{ZHAO2020100357}, UNet++~\cite{zhou2018unet++}, and UNet 3+~\cite{huang2020unet}) . The VAE-based models were trained in an unsupervised manner using unlabelled data, while the UNet-based models were trained in a supervised manner. Starting from their original 2D implementation, these networks were extended to accept 3D inputs of size $64^3$ by substituting all 2D layers with their 3D counterparts.

\subsubsection{Supervised models}
\label{sec:methods:supervised}
UNet is a popular encoder-decoder architecture that has shown promising results in many semantic voxel-wise classification tasks. MSS-UNet, UNet++, and UNet 3+ are extensions of the original UNet architecture. MSS-UNet incorporates multi-scale guidance in the decoding process during training, enabling it to capture more fine-grained details and to have a more coherent processing of information in the decoding stage. UNet++ includes a nested and dense skip-connection structure to capture more multi-scale features, while UNet 3+ uses a more powerful encoder with multi-resolution inputs.
MSS-UNet, UNet++, and UNet 3+ are deeply supervised models, which means they are trained with a loss function calculated on multiple inner layers to supervise the learning process effectively. In contrast, the original UNet architecture is not deeply supervised.

To ensure consistency, all UNet models were built using the same encoding/decoding building blocks, as in UNet++ and UNet 3+~\cite{huang2020unet}. This approach made it easier to compare the results of different architectures and understand how they impact the final outcome in voxel-wise classification tasks.

\subsubsection{Unsupervised models}
\label{sec:methods:unsupervised}
The VAE-based models were trained in an unsupervised manner to learn a compressed and disentangled representation of the input data. During training, the VAE models learned to reconstruct images from the compressed representations that resemble the input images as closely as possible. The reconstruction error, which quantifies the discrepancy between the input and output of the unsupervised models, was adopted as the anomaly score. Since the introduction of the VAE model in 2014 by Kingma and Welling, it has been used in a variety of studies for voxel-wise anomaly detection (e.g.~\cite{chen2019unsupervised, lin2020anomaly, chatterjee2021unsupervised}). The implemented VAE architecture follows the one described in a successive paper~\cite{zimmerer2018context}. The ceVAE model has identical architecture as VAE but a more complex definition of the loss. During training, ceVAE uses "masked" input data where certain patches within the image are fixed to a specific value. The model uses an ad-hoc loss function to infer the missing or distorted voxels within the masked zone, which helps the network to capture the context of the image. This peculiarity of the model may have a positive impact on the score, since it can prevent the network to learn to represent the pores within the training dataset. On the contrary, the gmVAE and vqVAE models are more complex than the VAE architecture, enabling them to catch features of the input 3D images that could not be interpreted by the coarser architecture of VAE. The gmVAE model assumes that each input data point's latent representation is generated by one of several possible Gaussian distributions, each with a different mean and variance, and identifies which distribution in the mixture is most likely to have generated the latent representation of each input data point during training. The vqVAE model is based on the idea of vector quantisation, where the continuous latent space is discretised into a set of discrete codes. The model comprises an encoder network that maps the input images to a discrete code book, followed by a decoder network that maps the discrete codes to the reconstructed input images. The vqVAE model was adapted to 3D inputs without additional alterations, except for an extra encoding/decoding stage that processes larger input patches of $64^3$ instead of the default $32^3$.

\subsection{Training}
\label{sec:methods:training}
The deep learning framework was based on the Pytorch~\cite{paszke2019pytorch} 2.0, Pytorch-lightning~\cite{falcon2020pytorchlightning} 1.9.0 and the CUDA~\cite{cuda} 11.7 libraries and it is publicly available\footnote{\url{https://github.com/snipdome/nn_3D-anomaly-detection}}. The 3D patch extraction, aggregation and data augmentation were based on the TorchIO libraries~\cite{perez2021torchio} version 0.18.84. A unique main seed propagated throughout the libraries ensures that all the extraction from random distributions were reproducible. Each of the models was trained with the Adam optimiser (learning rate of 0.0001) and halted through early stopping when the loss value did not decrease by more than 0.0001 for 40 consecutive epochs.

\subsection{Focal Tversky Loss function}
\label{sec:methods:FTL}
In our pore segmentation task, the number of voxels belonging to the
foreground class (pores) is much smaller than the number of voxels belonging to the background class, in the training dataset. This class imbalance results in a bias towards the background class during training, which leads to poor voxel-wise classification performance. In order to address the problem of class imbalance in semantic segmentation tasks, the Focal Tversky Loss (FTL) was proposed as a modification to the Tversky Loss~\cite{abraham2019novel}, and is defined as follows:
\begin{equation}
\text{FTL} = \left(1- \frac{\text{TP}}{\text{TP} + \alpha\, \text{FN} + \beta\, \text{FP}}\right)^\gamma
\label{eq:focal_tversky_loss}
\end{equation}
The FTL depends on the number of true negatives (TN), false negatives (FN), and false positives (FP), where FN and FP are weighted by $\alpha$ and $\beta$, respectively. By adjusting the values of these parameters, the FTL can be fine-tuned to emphasise either precision or recall. In addition, the FTL also includes a parameter $\gamma$, which controls the degree to which the FTL prioritises correcting misclassifications by adjusting the weight given to the Tversky Loss function. If $\gamma=1$, the FTL reduces to the standard Tversky loss and, if is also true that $\alpha=\beta=0.5$, to the Dice-S{\o}rensen loss. If $\gamma>1$, the FTL function will assign a higher weight to the correction of misclassifications. This means that the loss function will be more sensitive to false negatives and false positives, and the model will prioritise the correction of misclassifications over the correct classification of the majority class. As a result, the model will be better at identifying instances of the minority class but may struggle to accurately classify instances of the majority class. The degree to which the model's sensitivity to misclassifications increases will depend on the value of $\gamma$.
In case of deep supervision, the FTL is calculated at each supervised stage and averaged with geometric progression weights (1, 1/2, 1/4, etc.).

\subsection{Post-processing}
\label{sec:methods:post-process}
During the prediction or testing procedure, each of the models inferred patches belonging to the X-CT scan and then aggregated them back together to obtain an output volume with the same size as that of the input.

Only for the unsupervised networks, the output was post-processed to amend the scarce quality that these networks have in representing the fine details of the samples, as the surface. The surface of each of our samples has unique characteristics (due to different printing processes and polishing procedures), which can never be properly represented with an autoencoder (AE). While AEs are designed to learn a compressed minimal representation of the input, their ability to decode the latent space and reproduce high-fidelity input image representations depends on several factors, including the size and diversity of the training dataset, the complexity of the input data, and the architecture and hyperparameters of the model. As a result, the output can be blurry with lost or smoothed-out details. As this is a beneficial feature that makes the AEs potentially unable to reproduce anomalies that may be present in the training dataset, it comes with the cost of inaccuracies near the surface of the samples. This was compensated for by suppressing the anomaly score near the surface of samples. Specifically, the new voxel-wise anomaly score $A_{pores}$ was computed by subtracting from the former anomaly score $A$ the spatial (Gaussian) blurring of the absolute voxel-wise sum of derivatives of the predicted volume $\hat{V}$.

\begin{equation}
\label{eq:post-process}
A_{pores} = \max(0, A - \lambda  G_{\sigma}(\lVert \nabla_{\hat{V}} \rVert_1))
\end{equation}

The values for the standard deviation $\sigma$ of the Gaussian smearing kernel and the scaling factor $\lambda$ were chosen through an on-the-fly optimisation process of formula \ref{eq:opt_post-process}. The optimisation minimised the difference between the anomaly score and the Gaussian blurring of the absolute sum of derivatives of the inferred volume, using the mean of the L1-norm as a metric.

\begin{equation}
\label{eq:opt_post-process}
\lambda^*, \sigma^* = \arg \min_{\lambda, \sigma} \lVert A - \lambda G_{\sigma}(\lVert \nabla_{\hat{V}} \rVert_1) \rVert_1
\end{equation}

%
%
\section{Experiments}
\label{sec:experiments}
For the evaluations presented in this section, the labelled X-CT volumes were compared with the output of the DL models, after the output 3D patches were aggregated back together. More specifically,  the patch-extraction pipeline extracted overlapping patches from the input volume, each with half of their length overlapping with neighbouring patches. These patches were then classified by the neural networks and combined by computing an average value among the overlapping patches.

\subsection{Parameter search for the FTL function}
\label{sec:experiments:par-search}
As different values of the $\alpha, \beta, \gamma$ parameters sensibly affect the performance of models trained with the FTL function~\cite{iuso2022evaluation}, the optimal values were identified with grid search approach. A 5-fold cross-validation strategy evaluated the performance of the model with different parameter combinations, while the $\gamma$ parameter was kept at 0.5 (as in~\cite{iuso2022evaluation}). The grid search space spanned the parameter-space uniformly from 0.1 to 0.9 for each of the variables, for a total of 4 steps. For each combination of $\alpha$ and $\beta$, the model was trained in a 5-fold cross-validation, resulting in a total of 16 different combinations of $\alpha$ and $\beta$ and a total of 80 model trainings. In addition to the $\alpha$ and $\beta$ parameters, another grid search identified the optimal $\gamma$ parameter in the FTL. A higher value of $\gamma$ puts more emphasis on minimising false positives and false negatives, which can be useful in tasks where the cost of misclassification is high. So, even though the author of the FTL had suggested a value of $4/3$ for the $\gamma$ parameter~\cite{abraham2019novel}, the optimal $\gamma$ parameter turned out to vary for the current application of this work. The $\gamma$ grid search had a total of 8 steps ranging from $1/3$ to 2, for a total of 40 trainings.

\subsection{Cross-validation of performance of the DL models}
All the supervised and unsupervised models have been trained in a 5-fold cross-validation, for a total of 40 trainings. In the case of supervised models, they were trained with the optimal parameters found during the experiment \ref{sec:experiments:par-search}. After training, the performance has been evaluated, for each fold, on both the validation-set and the challenging test-set. 

\subsection{Cross-validation of performance of post-processed unsupervised models}
\label{sec:exp:post-process}
For this experiment, the unsupervised models are compared in cross-validation before and after the application of a post-processing algorithm presented in section \ref{sec:methods:post-process}. Since the post-processing happens after the aggregation of all the patches composing a X-CT volume, it is possible to compare the models before and after post-processing, without the need to re-train the models. Also in this case, the performance has been evaluated, for each fold, on both the validation-set and the challenging test-set.

\subsection{Cross-validation of performance of supervised models re-trained with unsupervised models}
\label{sec:exp:re-train}
In this experiment, the anomaly score of the (best performing) unsupervised model of experiment \ref{sec:exp:post-process} was used as label for the training of supervised models, for each fold. Training in such a way would make the overall pipeline unsupervised, which, apart from being a favourable feature for the user, it would theoretically allow the UNet-family to reproduce the task of the unsupervised model (and its post-processing algorithm). A total of 20 trainings has been performed.

\subsection{Cross-validation of performance of the best performing model in extreme visual scenarios}
\label{sec:exp:extreme-scenarios}
In this final experiment, the best performing model in the previous experiments has been tested when the image quality of the challenging test-set has been worsened by lowering X-ray exposure and number of projections. This test is designed to show how the performance decreases in extreme visual scenarios. The number of X-ray projections of the challenging test-set was reduced to 50\% and 33.3\%. The simulation of lower exposure of X-ray projections is achieved by adding Poisson distributed noise. The exposure was lowered to 75\%, 50\% and 25\% of the original values, which corresponded in an increase in the imaging noise over the X-ray projections.

%
%
\section{Results and discussions}
\label{sec:results}
Section \ref{sec:results:parameter-search} presents the cross-validation results for selecting the optimal parameters of the FTL function. These parameters were used to train all the supervised models employed in the voxel-wise classification task cross-validation, whose results are shown in sections \ref{sec:results:xval} and \ref{sec:results:unsup-labels}. Section \ref{sec:results:unsup-labels} compares the supervised models trained with the FTL function using heuristic labels and labels generated by the post-processed output of the best performing unsupervised model. The best performing unsupervised model was established based on the performance results presented in Section \ref{sec:results:post-process}.

\subsection{Parameter search for the FTL function}
\label{sec:results:parameter-search}
The initial parameter search for $\alpha$ and $\beta$ has been conducted on all the folds of the cross-validation, and the average results are shown in Table\,\ref{tab:ps-average}. As apparent from the results, the optimal values for the $\alpha$ and $\beta$ parameters are 0.633 and 0.1, respectively. Subsequently, with these optimal parameters, the optimal $\gamma$ parameter has been searched for each fold, and summary results are shown in Fig. \ref{tab:ps-gamma-average}. In this case, there is good agreement among folds that $\gamma = 1$ ensures the best performance. For the sake of completeness, the fold-wise results have been included in the appendix for both parameter searches (\ref{sec:appendix:ps}). 

\begin{table}
\centering
\renewcommand{\arraystretch}{1.8}
\begin{tabular}[*{6}c]{*{6}c} 
 & &\multicolumn{4}{c}{\textbf{Dice-S{\o}rensen score}}\\
\multirow{4}{*}{$\mathrm{\alpha}$} & 0.9 &{\color{white}\grada{0.74}$\mathrm{\pm}$ 0.06}&{\color{white}\grada{0.74}$\mathrm{\pm}$ 0.04}& {\color{white}\grada{0.75}$\mathrm{\pm}$ 0.07}&{\color{white}\grada{0.75}$\mathrm{\pm}$ 0.07}\\
 & 0.63 &{\color{white}\grada{0.78}$\mathrm{\pm}$ 0.04}&{\color{white}\grada{0.76}$\mathrm{\pm}$ 0.06}&{\color{white}\grada{0.76}$\mathrm{\pm}$ 0.06}&{\color{white}\grada{0.73}$\mathrm{\pm}$ 0.07}\\
 & 0.37 &{\color{white}\grada{0.75}$\mathrm{\pm}$ 0.05}&{\color{white}\grada{0.76}$\mathrm{\pm}$ 0.07}&{\color{white}\grada{0.74}$\mathrm{\pm}$ 0.07}&{\color{white}\grada{0.70}$\mathrm{\pm}$ 0.05}\\
 & 0.1   &{\color{white}\grada{0.76}$\mathrm{\pm}$ 0.06}&{\color{white}\grada{0.73}$\mathrm{\pm}$ 0.07}&{\color{white}\grada{0.72}$\mathrm{\pm}$ 0.07}&{\color{white}\grada{0.71}$\mathrm{\pm}$ 0.06}\\
 & & 0.1 & 0.37 & 0.63 & 0.9\\
 & &\multicolumn{4}{c}{$\mathrm{\beta}$}\\
\end{tabular}
\caption{\rule{0pt}{3ex}Average Dice-Sørensen score and standard error of the models evaluated across the related validation dataset, depending on the $\alpha$/$\beta$ parameters of the FTL.}
\label{tab:ps-average}
\end{table}

\begin{table}
\centering
\renewcommand{\arraystretch}{1.5}
\setlength{\tabcolsep}{3pt}
\begin{tabular}[c c c c c c c c c]{c c c c c c c c c}
&\multicolumn{8}{c}{\textbf{Dice-S{\o}rensen score}}\\
&\footnotesize{\color{white}\gradb{0.76}$\mathrm{\pm}$0.04}&\footnotesize{\color{white}\gradb{0.78}$\mathrm{\pm}$0.04}&\footnotesize{\color{white}\gradb{0.79}$\mathrm{\pm}$0.03}&\footnotesize{\color{white}\gradb{0.82}$\mathrm{\pm}$0.04}&\footnotesize{\color{white}\gradb{0.72}$\mathrm{\pm}$0.03}&\footnotesize{\color{white}\gradb{0.70}$\mathrm{\pm}$0.06}&\footnotesize{\color{white}\gradb{0.64}$\mathrm{\pm}$0.06}&\footnotesize{\color{white}\gradb{0.59}$\mathrm{\pm}$0.04}\\
$\footnotesize{\mathrm{\gamma}}$&0.33&0.5&0.67&1.0&1.33&1.5&1.67&2.0\\
\end{tabular}
\caption{\rule{0pt}{3ex}Average Dice-Sørensen score  and standard error of the models evaluated across the related validation dataset, depending on the $\gamma$ parameters of the FTL.}
\label{tab:ps-gamma-average}
\end{table}

\subsection{Cross-validation of performance of the DL models}
\label{sec:results:xval}
The segmentation results of the cross-validation technique were evaluated using two metrics: the area under the ROC curve (AUC) and the average precision (AP) of the precision-recall (PR) curve. While the AUC is a commonly used metric, it can be misleading in the presence of class imbalance~\cite{hanczar2010small, saito2015precision}. To address this issue, PR curves were used to evaluate the performance of algorithms, as recommended by~\cite{saito2015precision}. Therefore, both PR and ROC curves were used to evaluate the models.

\begin{table}
\centering
\begin{tabular}{ccc}
Model & AP & AUC\\
\hline\hline
MSS-UNet$^{\mathrm{\triangle}}$           &  0.784 $\pm$ 0.050  &  0.975 $\pm$ 0.013\\
UNet$^{\mathrm{\triangle}}$               &  0.815 $\pm$ 0.025 &  0.983 $\pm$ 0.009\\
UNet++$^{\mathrm{\triangle}}$             &  0.750 $\pm$ 0.026 &  0.974 $\pm$ 0.009\\
UNet-3+$^{\mathrm{\triangle}}$            &  \textbf{0.876 $\pm$ 0.036} &  0.992 $\pm$ 0.003\\
\hline
VAE$^{\mathrm{\diamondsuit}}$                &  0.728 $\pm$ 0.095 &  0.999 $\pm$ 0.001\\
ceVAE$^{\mathrm{\diamondsuit}}$              &  \textbf{0.746 $\pm$ 0.094} &  0.999 $\pm$ 0.001\\
gmVAE$^{\mathrm{\diamondsuit}}$              &  0.607 $\pm$ 0.156 &  0.974 $\pm$ 0.014\\
vqVAE$^{\mathrm{\diamondsuit}}$              &  0.602 $\pm$ 0.128 &  0.990 $\pm$ 0.004\\
\hline
\end{tabular}
\caption{\rule{0pt}{3ex}Average ROC-AUC and AP scores (with confidence interval) of the supervised ($\mathrm{\triangle}$) and unsupervised ($\mathrm{\diamondsuit}$) models evaluated on the validation dataset.}
\label{tab:AUC-AP:validation}
\end{table}

\begin{table}
\centering
\begin{tabular}{ccc}
Model & AP & AUC\\
\hline\hline
MSS-UNet$^{\mathrm{\triangle}}$           &  0.572 $\pm$ 0.036  &  0.855 $\pm$ 0.017\\
UNet$^{\mathrm{\triangle}}$               &  0.581 $\pm$ 0.020 &  0.880 $\pm$ 0.014\\
UNet++$^{\mathrm{\triangle}}$             &  \textbf{0.583 $\pm$ 0.036} &  0.848 $\pm$ 0.014\\
UNet-3+$^{\mathrm{\triangle}}$            &  0.541 $\pm$ 0.008 &  0.882 $\pm$ 0.008\\
\hline
VAE$^{\mathrm{\diamondsuit}}$                &  0.615 $\pm$ 0.038 &  0.990 $\pm$ 0.002\\
ceVAE$^{\mathrm{\diamondsuit}}$              &  \textbf{0.635 $\pm$ 0.021} &  0.990 $\pm$ 0.001\\
gmVAE$^{\mathrm{\diamondsuit}}$              &  0.274 $\pm$ 0.092 &  0.838 $\pm$ 0.044\\
vqVAE$^{\mathrm{\diamondsuit}}$              &  0.313 $\pm$ 0.025 &  0.871 $\pm$ 0.009\\
\hline
\end{tabular}
\caption{\rule{0pt}{3ex}Average ROC-AUC and AP (with confidence interval) of the supervised ($\mathrm{\triangle}$) and unsupervised ($\mathrm{\diamondsuit}$) models evaluated on the challenging test-set.}
\label{tab:AUC-AP:challenge}
\end{table}

\begin{figure*}
\centering
\begin{subfigure}{.498\textwidth}
  \includegraphics[width=\linewidth]{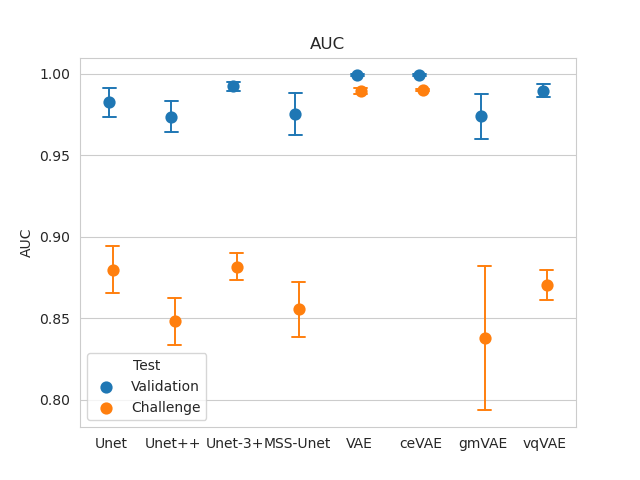}
  \label{fig:summary:nets:AUC}
\end{subfigure}%
\begin{subfigure}{.498\textwidth}
  \includegraphics[width=\linewidth]{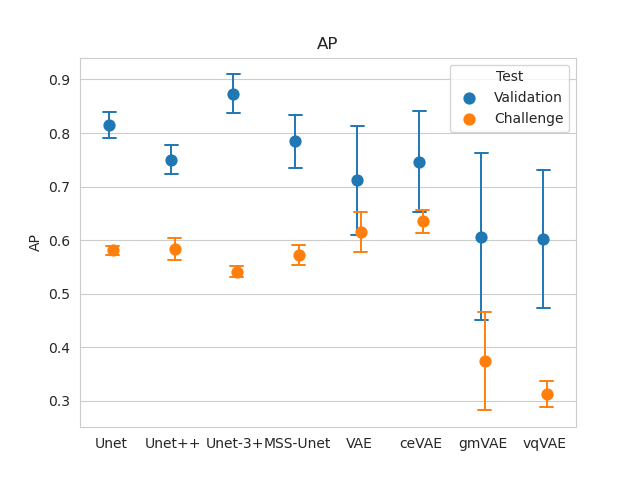}
  \label{fig:summary:nets:AP}
\end{subfigure}%
\caption{Point-plots of the average ROC-AUC and AP scores (with confidence interval) of the models evaluated on the validation dataset and on the challenging dataset. The quantitative values are shown in Table \ref{tab:AUC-AP:validation} and Table \ref{tab:AUC-AP:challenge}.}
\label{fig:summary:nets}
\end{figure*}

The voxel-wise classification task of the models was evaluated for each fold, whose summary ROC-AUC and AP values are shown in Fig. \ref{fig:summary:nets} for the validation dataset and the challenging test set. The cross-validated results related to the validation dataset (represented with blue colour in Fig. \ref{fig:summary:nets}) indicate that supervised models have been generally better trained to be consistent with labels than the unsupervised methods. The results on the challenging dataset with high artefacts and manually labelled (represented with orange colour) show a clear drop of the score for all the models, as expected for the considerations in \ref{sec:methods:datasets}. Moreover, it is noticeable that the score of some of the unsupervised models is even higher than that of the supervised ones for the challenging dataset. Although these results may not seem consistent with the validation dataset, it should be noted that in both cases the labels were generated in different ways: either with a heuristic labelling algorithm or via manual annotation. Among the supervised models, there is no significant difference in performance, which suggests that deep supervision and the different architecture of the models is not inducing a significant difference in performance. On the other hand, a noticeable difference in scores is present between ceVAE and gmVAE/vqVAE on the challenging dataset, which is significant for vqVAE with a confidence of 95\% (Welch's t-test, p-value $1.98 * 10^{-4}$ (AUC) and $1.05 * 10^{-5}$ (AP)). The higher degree of complexity of gmVAE and vqVAE is not favourable to the segmentation task by the mean of the anomaly score. These models have been capable of learning how to reproduce defects within the input samples, so the reconstruction error is not as high in the proximity of defects as it is with simpler VAEs. On another note, VAE and ceVAE are most robust with respect to the quality of the input image, since the AP/AUC scores are almost unvaried between the validation and the challenging test-set (AP/AUC differences lower than or approximately equal to a decimal point), when opposed to the other models (AP/AUC differences exceeding a decimal point).

\begin{figure*}
\centering
\begin{subfigure}{.498\textwidth}
  \includegraphics[width=\linewidth]{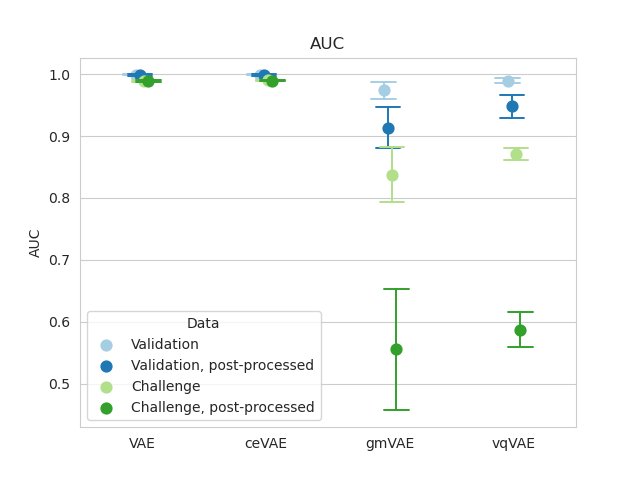}
  \label{fig:summary:post-process:AUC}
\end{subfigure}%
\begin{subfigure}{.498\textwidth}
  \includegraphics[width=\linewidth]{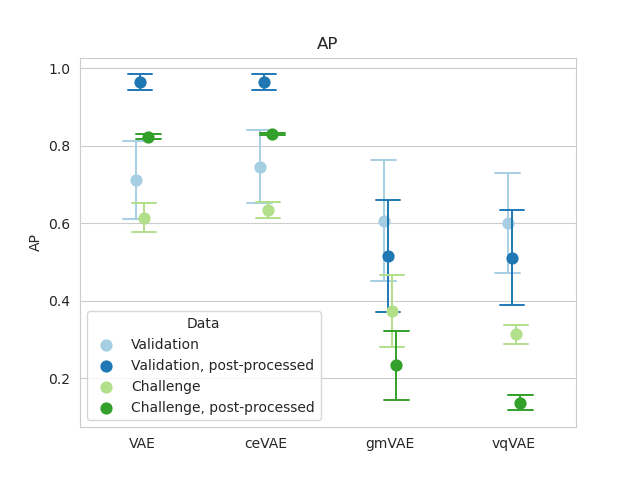}
  \label{fig:summary:post-process:AP}
\end{subfigure}%
\caption{Point-plots of the average ROC-AUC and AP of the models (with confidence interval) evaluated on the validation dataset and on the challenging dataset, with and without post-processing. Solely the performance of unsupervised models is shown, since the post-processing of the output is defined for them only. The values in textual form are shown in Table \ref{tab:AUC-AP:post-process:validation} and Table \ref{tab:AUC-AP:post-process:challenge}.}
\label{fig:summary:post-process}
\end{figure*}

\begin{table}
\centering
\begin{tabular}{ccc}
Model & AP & AUC\\
\hline\hline
VAE                &  0.964 $\pm$ 0.020 &  0.998 $\pm$ 0.001\\
ceVAE              &  0.964 $\pm$ 0.019 &  0.999 $\pm$ 0.001\\
gmVAE              &  0.516 $\pm$ 0.144 &  0.913 $\pm$ 0.033\\
vqVAE              &  0.512 $\pm$ 0.122 &  0.948 $\pm$ 0.019\\
\hline
\end{tabular}
\caption{\rule{0pt}{3ex}Average ROC-AUC and AP (with confidence interval) of the unsupervised models evaluated on the validation dataset, with post-processing of the output. Solely the performance of unsupervised models is shown, since the post-processing of the output is defined for them only.}
\label{tab:AUC-AP:post-process:validation}
\end{table}

\begin{table}
\centering
\begin{tabular}{ccc}
Model & AP & AUC\\
\hline\hline
VAE                &  0.824 $\pm$ 0.007 &  0.989 $\pm$ 0.001\\
ceVAE              &  \textbf{0.830 $\pm$ 0.003} &  0.989 $\pm$ 0.001\\
gmVAE              &  0.234 $\pm$ 0.089 &  0.555 $\pm$ 0.099\\
vqVAE              &  0.138 $\pm$ 0.020 &  0.587 $\pm$ 0.028\\
\hline
\end{tabular}
\caption{\rule{0pt}{3ex}Average ROC-AUC and AP (with confidence interval) of the unsupervised models evaluated on the challenging test set, with post-processing of the output. Solely the performance of unsupervised models is shown, since the post-processing of the output is defined for them only.}
\label{tab:AUC-AP:post-process:challenge}
\end{table}

\subsection{Cross-validation of performance of post-processed unsupervised models}
\label{sec:results:post-process}
\begin{figure}
\centering
\begin{subfigure}{.244\textwidth}
  \centering
  \includegraphics[width=\linewidth]{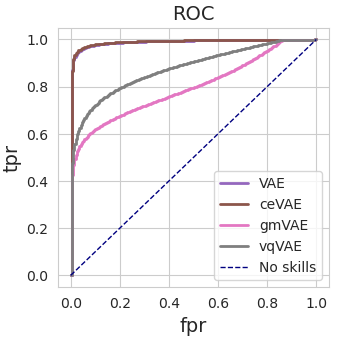}
\end{subfigure}
\begin{subfigure}{.244\textwidth}
  \centering
  \includegraphics[width=\linewidth]{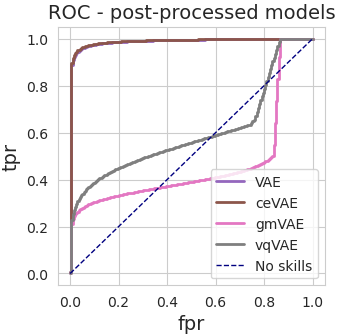}
\end{subfigure}
\begin{subfigure}{.244\textwidth}
  \centering
  \includegraphics[width=\linewidth]{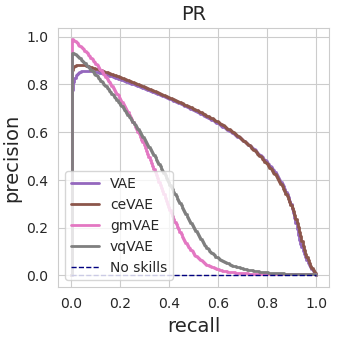}
\end{subfigure}
\begin{subfigure}{.244\textwidth}
  \centering
  \includegraphics[width=\linewidth]{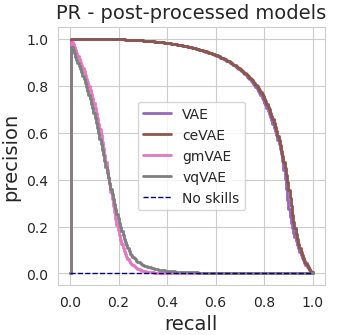}
\end{subfigure}
\caption{Graph of the ROC and PR curves of cross-validated performance for all models. The graphs represent the median trend of the fold-wise performance on the challenging dataset without (left) and with post-processing (right) of the aggregated output.}
\label{fig:graphs:normal-vs-postprocess}
\end{figure}

By applying post-processing to the output of the VAE models (Fig. \ref{fig:summary:post-process}), the considerations of the previous section about supervised models become more evident. When post-processing is applied to the output of the VAE and ceVAE models, which have not learned to visually represent pores, their AP scores increase by almost 2 decimal points on both datasets, while their AUC remains almost unchanged. On the other hand, post-processing adversely affected the performance of gmVAE and vqVAE, which is to be expected since the derivative of the output of these models is non-negligible near the edge of the sample as well as near the pores. This behaviour is noticeable in the ROC and PR classifier curves for the challenging case as shown in Fig. \ref{fig:graphs:normal-vs-postprocess} (other ROC and PR graphs are shown in the \ref{sec:appendix:graphs}). The more complex nature of gmVAE/vqVAE models has allowed these models to reproduce the defects within the samples, which leads to a reduction of the anomaly score, so compromising the performance. This effect becomes even more pronounced when post-processing is applied. It is noticeable from these results that having a more complex architecture, is not always desirable, especially when anomalies are present within the training dataset. Additionally, it can be observed from Fig. \ref{fig:summary:nets} and Fig. \ref{fig:summary:post-process} that the scores of VAE and ceVAE are still resilient against the poor image quality of the challenging test-set, compared to the drastic drop in performance of the supervised networks.

\subsection{Cross-validation of supervised models trained with labels generated by an unsupervised model}
\label{sec:results:unsup-labels}

\begin{table}
\centering
\begin{tabular}{ccc}
Model & AP & AUC\\
\hline\hline
MSS-UNet           &  0.651 $\pm$ 0.008  &  0.889 $\pm$ 0.005\\
UNet               &  0.639 $\pm$ 0.008  &  0.882 $\pm$ 0.004\\
UNet++             &  \textbf{0.751 $\pm$ 0.030} &  0.902 $\pm$ 0.015\\
UNet-3+            &  0.627 $\pm$ 0.006 &  0.894 $\pm$ 0.006\\
\hline
\end{tabular}
\caption{\rule{0pt}{3ex}Average ROC-AUC and AP (with confidence interval) of the supervised models re-trained with the labels generated by ceVAE and evaluated on the challenging dataset.}
\label{tab:AUC-AP:retrain:challenge}
\end{table}

\begin{figure}
\centering
\begin{subfigure}{.245\textwidth}
  \centering
  \includegraphics[width=\linewidth]{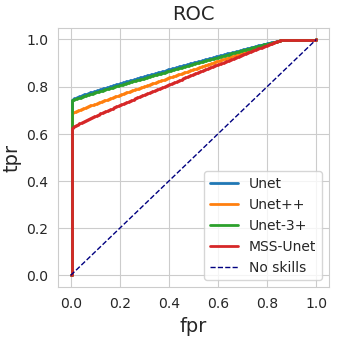}
\end{subfigure}%
\begin{subfigure}{.245\textwidth}
  \centering
  \includegraphics[width=\linewidth]{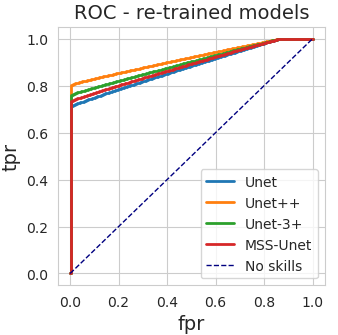}
\end{subfigure}
\begin{subfigure}{.245\textwidth}
  \centering
  \includegraphics[width=\linewidth]{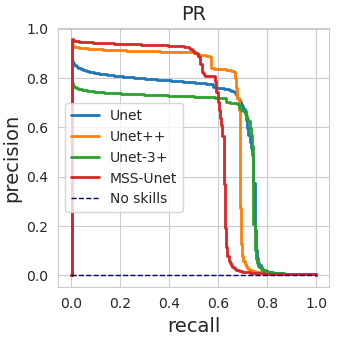}
\end{subfigure}
\begin{subfigure}{.245\textwidth}
  \centering
  \includegraphics[width=\linewidth]{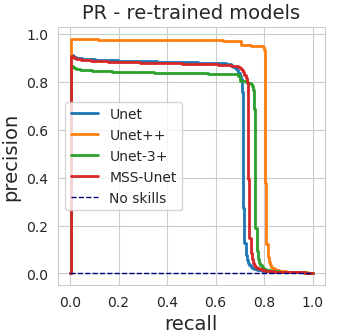}
\end{subfigure}
\caption{Graph of the ROC and PR curves of cross-validated performance for all models. The graphs represent the median trend of the fold-wise performance on the challenging dataset, with Otsu-based labels (left) and post-processed ceVAE-generated labels (right).}
\label{fig:graphs:normal-vs-retrain}
\end{figure}

\begin{figure*}
\centering
\begin{subfigure}{.498\textwidth}
  \includegraphics[width=\linewidth]{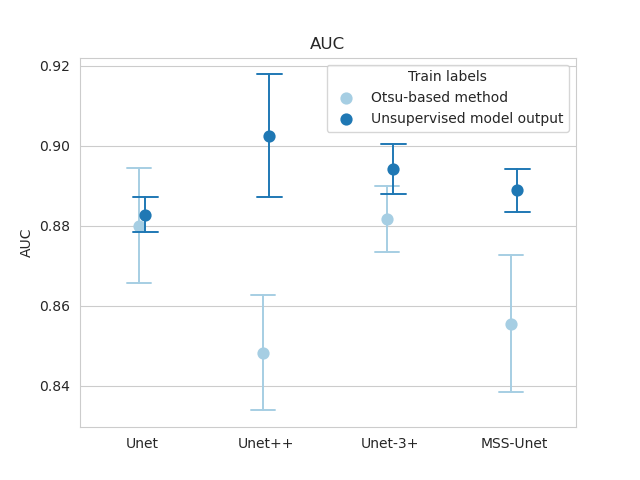}
  \label{fig:summary:train-labels:AUC}
\end{subfigure}%
\begin{subfigure}{.498\textwidth}
  \includegraphics[width=\linewidth]{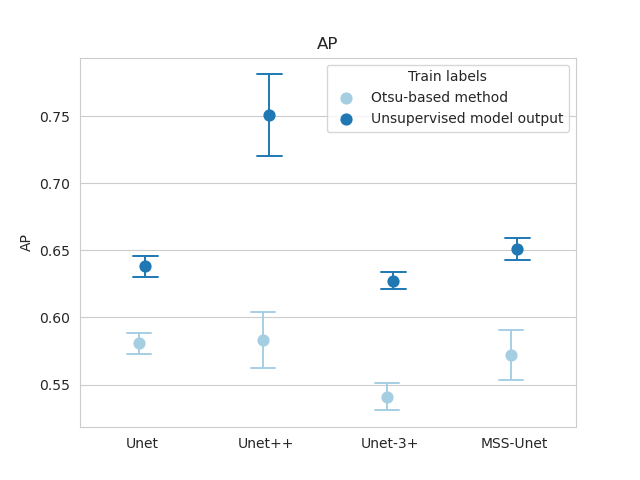}
  \label{fig:summary:train-labels:AP}
\end{subfigure}%
\caption{Point-plots of the average ROC-AUC and AP (with confidence interval) of the supervised models evaluated on the challenging dataset. The graphs highlight the different performance when these models were supervised by the Otsu-based method and with the labels provided by the unsupervised models. The values in textual form are shown in Table \ref{tab:AUC-AP:retrain:challenge}}
\label{fig:summary:train-labels}
\end{figure*}

By using ceVAE (the best performing model) to generate labels for the samples, the supervised models could be trained from scratch to detect pores. The necessary steps for the production of these labels by ceVAE were the post-processing (with the algorithm described in section \ref{sec:methods:datasets}) and the suppression of smaller pores. The results are shown in Fig. \ref{fig:summary:train-labels} and Fig. \ref{fig:graphs:normal-vs-retrain}. Higher performance is achieved by using the unsupervised labels, confirmed by both AUC and AP for all the models. These results confirm the observations in section \ref{sec:results:xval} that the different architectures of the models are not significantly affecting the scores for this voxel-wise classification task.

\subsection{Cross-validation of performance of the best performing model in extreme visual scenarios}
\label{sec:results:extreme-scenarios}
\begin{figure*}
\centering
\begin{subfigure}{.498\textwidth}
  \includegraphics[width=\linewidth]{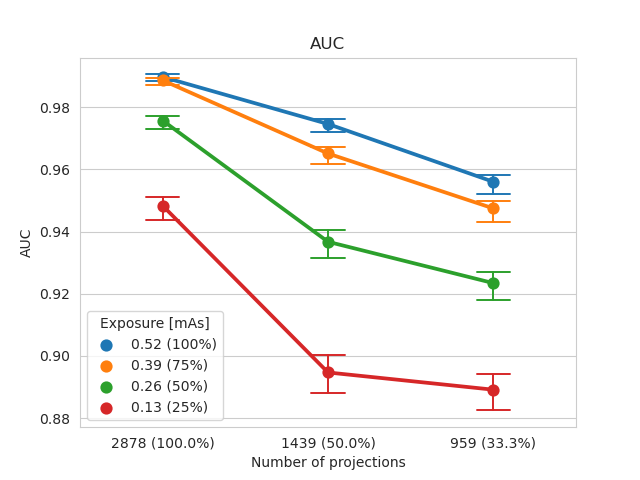}
  \label{fig:summary:extreme-scenarios:AUC}
\end{subfigure}%
\begin{subfigure}{.498\textwidth}
  \includegraphics[width=\linewidth]{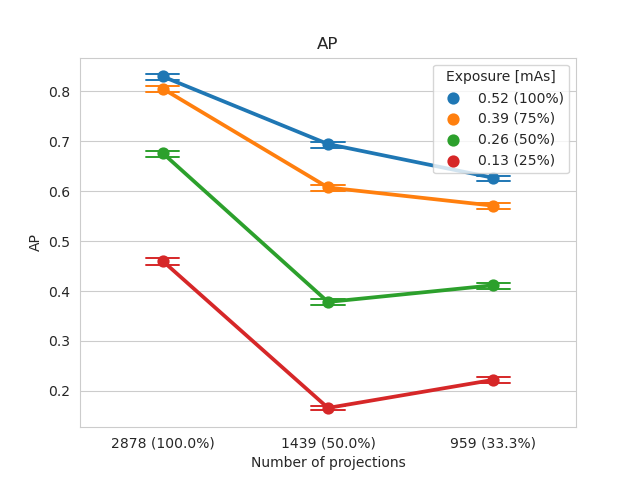}
  \label{fig:summary:extreme-scenarios:AP}
\end{subfigure}%
\caption{Point-plots of the average ROC-AUC and AP (with confidence interval) of the anomaly score of ceVAE evaluated on the challenging test-set when the image quality is lowered by reducing the number of X-ray projections or exposure.}
\label{fig:summary:extreme-scenarios}
\end{figure*}

By reducing the number of X-ray projections of the challenging X-CT scan and reducing the exposure of each X-ray projection, the quality of the reconstructed X-CT scan decreased. The best performing model, which was shown to be the post-processed ceVAE, was applied to these X-CT scans. An exemplary visual representation of the voxel-wise classification is shown in Fig. \ref{fig:res:extreme-scenarios}, related to the post-processed output of the ceVAE model, trained on the 1st fold. In this figure, a small portion of a slice of the cube is shown, in which pores are visible that were induced with off-nominal parameters of the melting laser during the printing. The degradation of the classification performance is noticeable due to the increasing number of voxels classified as pores (as shown in Fig. \ref{fig:summary:extreme-scenarios}). Interestingly, while reducing the number of X-ray projections from 4283 (the dataset used for training/validation) to 2878 (the original challenging test-set) did not significantly affect the performance (Fig. \ref{fig:summary:post-process}), further reductions in the number of projections had a significant impact on the performance scores (Fig. \ref{fig:summary:extreme-scenarios}). Another point to note is the trend exhibited by the AP scores at low exposure levels ranging from 50-25\%. Specifically, reducing the number of projections from 50\% to 33.3\% led to a slight increase in the AP scores. When data is highly noisy and the number of projections is relatively low, adding some more X-ray projections may not always lead to better image quality of the reconstructed X-CT scans. This is because the additional (noisy) projections can also introduce more noise into the reconstructed images. This can be observed from the fact that the trend gradually disappears as the exposure level increases from 25\% to 100\%.

\pgfmathparse{1/6.3} 
\begin{table} 
\setlength\tabcolsep{0.pt}
\begin{tabular}{ccccccc} 
&\multicolumn{2}{c}{2878 projections (100\%)}&\multicolumn{2}{c}{1439 projections (50\%)}&\multicolumn{2}{c}{959 projections (33.3\%)}\\
\rotatebox{90}{  0.52 mAs (100\%)}\,&\includegraphics[width=\pgfmathresult\columnwidth]{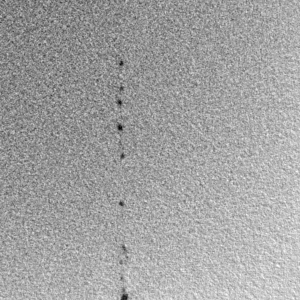}&
\includegraphics[width=\pgfmathresult\columnwidth]{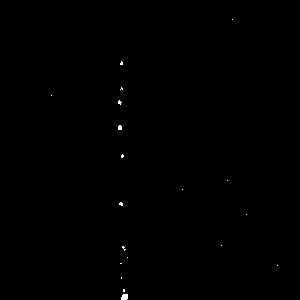}&\hspace{1.pt}
\includegraphics[width=\pgfmathresult\columnwidth]{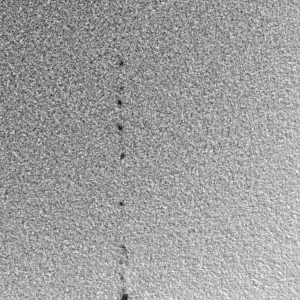}&
\includegraphics[width=\pgfmathresult\columnwidth]{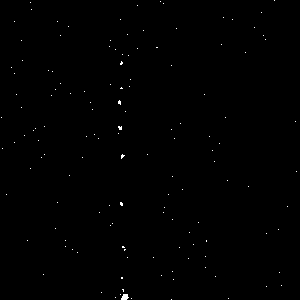}&\hspace{1.pt}
\includegraphics[width=\pgfmathresult\columnwidth]{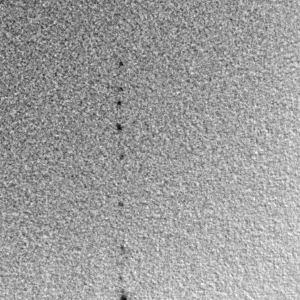}&
\includegraphics[width=\pgfmathresult\columnwidth]{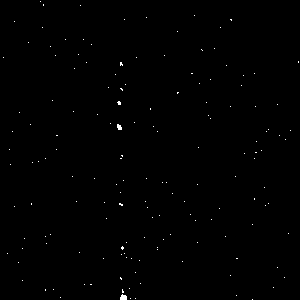}\\
\rotatebox{90}{  0.39 mAs (75\%)}\,&\includegraphics[width=\pgfmathresult\columnwidth]{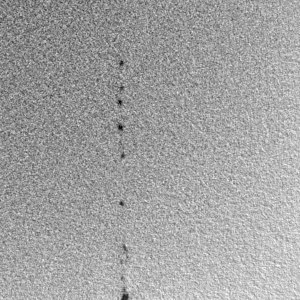}&
\includegraphics[width=\pgfmathresult\columnwidth]{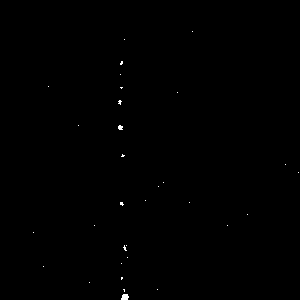}&\hspace{1.pt}
\includegraphics[width=\pgfmathresult\columnwidth]{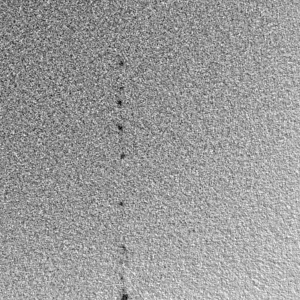}&
\includegraphics[width=\pgfmathresult\columnwidth]{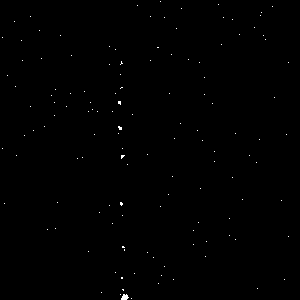}&\hspace{1.pt}
\includegraphics[width=\pgfmathresult\columnwidth]{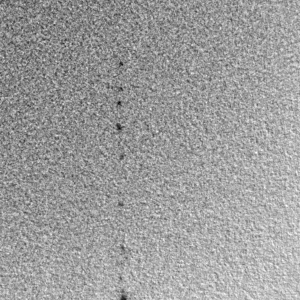}&
\includegraphics[width=\pgfmathresult\columnwidth]{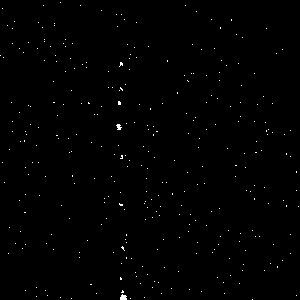}\\
\rotatebox{90}{  0.26 mAs (50\%)}\,&\includegraphics[width=\pgfmathresult\columnwidth]{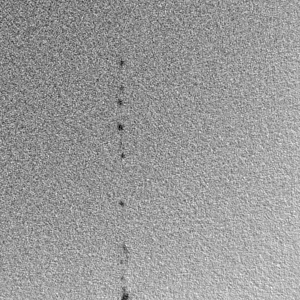}&
\includegraphics[width=\pgfmathresult\columnwidth]{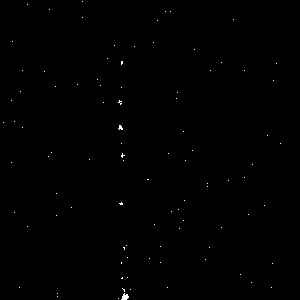}&\hspace{1.pt}
\includegraphics[width=\pgfmathresult\columnwidth]{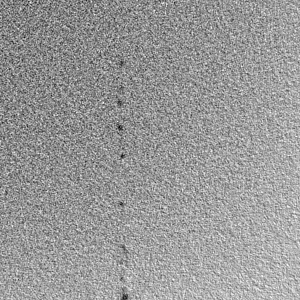}&
\includegraphics[width=\pgfmathresult\columnwidth]{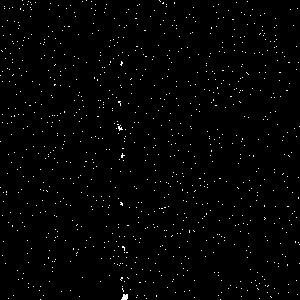}&\hspace{1.pt}
\includegraphics[width=\pgfmathresult\columnwidth]{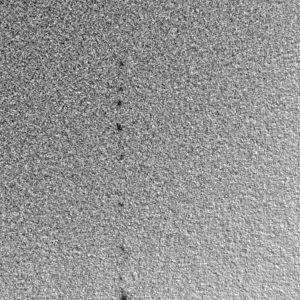}&
\includegraphics[width=\pgfmathresult\columnwidth]{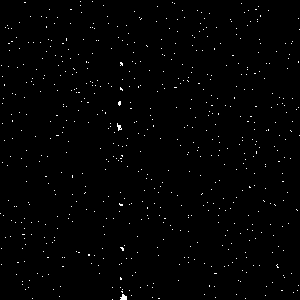}\\
\rotatebox{90}{  0.13 mAs (25\%)}\,&\includegraphics[width=\pgfmathresult\columnwidth]{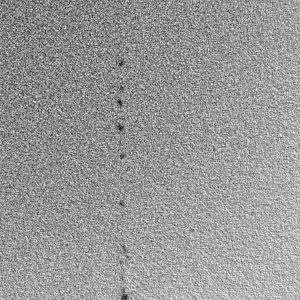}&
\includegraphics[width=\pgfmathresult\columnwidth]{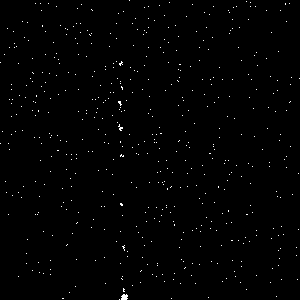}&\hspace{1.pt}
\includegraphics[width=\pgfmathresult\columnwidth]{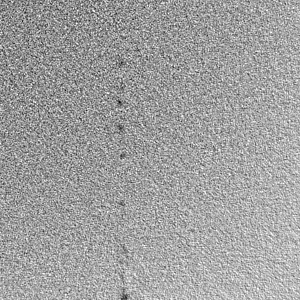}&
\includegraphics[width=\pgfmathresult\columnwidth]{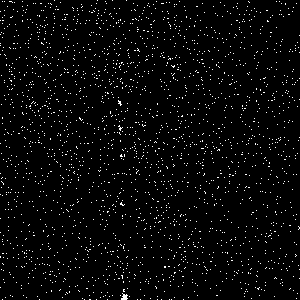}&\hspace{1.pt}
\includegraphics[width=\pgfmathresult\columnwidth]{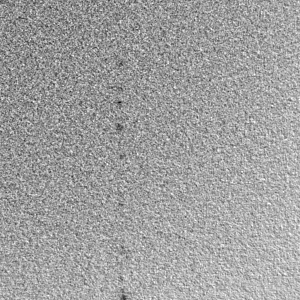}&
\includegraphics[width=\pgfmathresult\columnwidth]{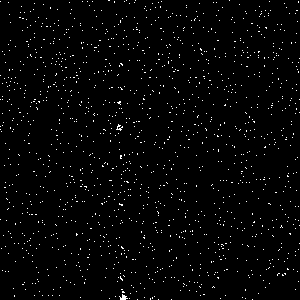}
\end{tabular} 
\captionof{figure}{A portion of a X-CT slice is shown in each row and column by modifying the number of X-ray projections and exposure of each X-ray projection. Each input slice is shown together with the label mask predicted by ceVAE (trained on the 1st fold). The degradation of the classification performance is noticeable from the raising number of voxels that are classified as pores (white colour in the predicted mask).} 
\label{fig:res:extreme-scenarios} 
\end{table}

%
%
\section{Conclusions}
With a 3D patch-based approach, we extended several models (UNet, UNet++, UNet 3+, MSS-UNet, VAE, ceVAE, gmVAE, vqVAE) from their original 2D implementation to 3D, and trained them for the voxel-wise classification of pores in AM samples. These models were trained in a supervised (UNet-family of models) and unsupervised (VAE-family of models) way. For the unsupervised models, a post-processing technique was applied to reduce the anomaly score at the surface of the samples, which is shown to be beneficial for VAE and ceVAE in terms of AP, but harmful for gmVAE and vqVAE. The more complex nature of the gmVAE/vqVAE models allowed these models to reproduce the defects within the training samples, compromising the performance. The performance of the resulting VAE/ceVAE models was demonstrated to be quite resilient to lower image quality, which was not the case for the supervised models. Training supervised models with labels from the best unsupervised model (ceVAE) improved their performance but did not surpass the unsupervised model. Deep supervision among the supervised models did not consistently impact performance. The study suggests that ceVAE captured the statistical properties of 3D patches of the input dataset in a more robust way than the UNet family. This particular finding is evidenced by the unmatched performance of ceVAE in the segmentation task of the challenging dataset shown here and confirms similar findings in anomaly detection in MRI images~\cite{chatterjee2022strega}. While the focus of our research revolved around porosity analysis in the AM process, it opens up possibilities for broader anomaly detection in AM samples, such as identifying impurities in printing materials, microstructural inhomogeneities, or the loss of alloying elements due to vaporisation. 

\backmatter

\section*{Acknowledgements}
This study is financially supported by the imec ICON project Multiplicity, the Research Foundation Flanders (FWO, SBO grant no. S007219N) and the Flemish Government under the “Onderzoeksprogramma Artificiele Intelligentie (AI) Vlaanderen” programme.

\section*{Statements and Declarations}
\subsection*{Data Availability}
The data presented in this study are available upon reasonable request from the corresponding author. 

\subsection*{Author contributions}
Domenico Iuso devised the project, the main conceptual ideas and worked out all of the technical details. Soumick Chatterjee contributed to the conceptual ideas and the design of the research. Sven Cornelissen and Dries Verhees contributed with the design and printing of AM samples. Jan De Beenhouwer and Jan Sijbers supervised the project. All authors discussed the results, commented on the manuscript and approved the final manuscript.

\subsection*{Compliance with Ethical Standards}
The authors have no competing interests to declare that are relevant to the content of this article.

\bibliography{main.bib}

\begin{appendices}

\section{Classifier graphs for the voxel-wise classification task}
\label{sec:appendix:graphs}
\begin{figure}
\centering
\begin{subfigure}{.245\textwidth}
  \centering
  \includegraphics[width=\linewidth]{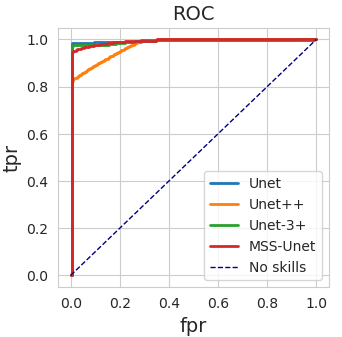}
\end{subfigure}%
\begin{subfigure}{.245\textwidth}
  \centering
  \includegraphics[width=\linewidth]{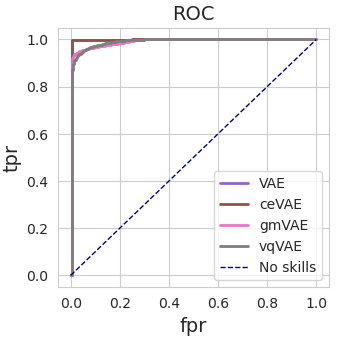}
\end{subfigure}
\begin{subfigure}{.245\textwidth}
  \centering
  \includegraphics[width=\linewidth]{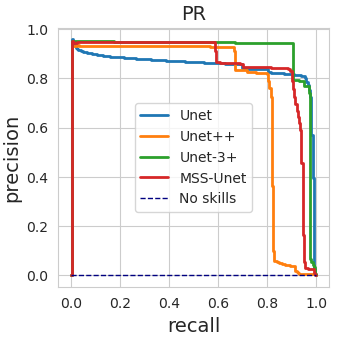}
\end{subfigure}
\begin{subfigure}{.245\textwidth}
  \centering
  \includegraphics[width=\linewidth]{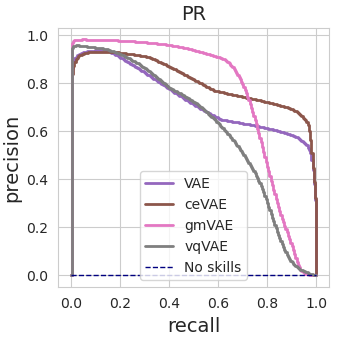}
\end{subfigure}
\caption{Graph of the ROC and PR curves of cross-validated performance for all models. The graphs represent the median trend of the fold-wise performance on related validation dataset.}
\label{fig:graphs:xval}
\end{figure}

\begin{figure}
\centering
\begin{subfigure}{.245\textwidth}
  \centering
  \includegraphics[width=\linewidth]{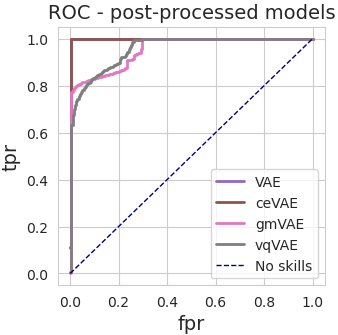}
\end{subfigure}
\begin{subfigure}{.245\textwidth}
  \centering
  \includegraphics[width=\linewidth]{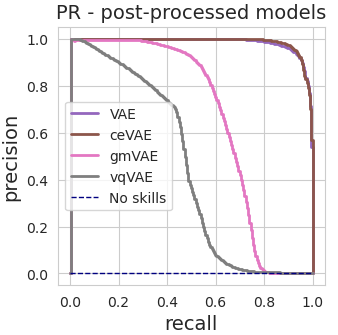}
\end{subfigure}
\caption{Graph of the ROC and PR curves of cross-validated performance for the unsupervised models. The graphs represent the median trend of the fold-wise performance on the related validation dataset, when the output of the models is post processed.}
\label{fig:graphs:post-process}
\end{figure}

The ROC and PR graphs of the voxel-wise classification results that were not shown in previous sections are reported here. In \reffig{fig:graphs:xval}, there are the performance graphs of supervised and unsupervised models evaluated on the related validation dataset. The graphs are aligned with the findings discussed in section \ref{sec:results:xval} and \ref{sec:results:unsup-labels}. In \reffig{fig:graphs:post-process} are shown the performance of the unsupervised models only, since they show the classification scores of the post-processed output. The scores were obtained from the fold-wise performance on the related validation dataset, where is noticeable an increase of performance for VAE/ceVAE and a decrease for gmVAE/vqVAE if compared with \reffig{fig:graphs:xval} (right), in accordance with the findings in section \ref{sec:results:unsup-labels}.

\section{Cross-validation graphs for the FTL parameter search per each fold}
\label{sec:appendix:ps}
For each of the 5 folds of the cross-validation, there is a total of 16 trainings for the $\alpha$/$\beta$ parameter, which are presented in Table \ref{tab:ps-folds}. For the $\gamma$ parameter, there is a total of 8 trainings per fold and the values of the Dice-Sørensen are shown in Table \ref{tab:ps-gamma-folds}.

\begin{table}
    \begin{minipage}{.5\linewidth}
\centering
\renewcommand{\arraystretch}{1.5}
\begin{tabular}[*{6}c]{*{6}c} 
 & &\multicolumn{4}{c}{\textbf{Dice-S{\o}rensen score - Fold 1}}\\
\multirow{4}{*}{$\mathrm{\alpha}$} & 0.9 &{\color{white}\gradda{0.82}}&{\color{white}\gradda{0.81}}& {\color{white}\gradda{0.97}}&{\color{white}\gradda{0.97}}\\
 & 0.63 &{\color{white}\gradda{0.86}}&{\color{white}\gradda{0.97}}&{\color{white}\gradda{0.97}}&{\color{white}\gradda{0.96}}\\
 & 0.37 &{\color{white}\gradda{0.96}}&{\color{white}\gradda{0.97}}&{\color{white}\gradda{0.96}}&{\color{white}\gradda{0.83}}\\
 & 0.1  &{\color{white}\gradda{0.97}}&{\color{white}\gradda{0.95}}&{\color{white}\gradda{0.94}}&{\color{white}\gradda{0.84}}\\
 & & 0.1 & 0.37 & 0.63 & 0.9\\
 & &\multicolumn{4}{c}{$\mathrm{\beta}$}\\
\end{tabular}
    \end{minipage}%
    \begin{minipage}{.5\linewidth}
\centering
\renewcommand{\arraystretch}{1.5}
\begin{tabular}[*{6}c]{*{6}c} 
 & &\multicolumn{4}{c}{\textbf{Dice-S{\o}rensen score - Fold 2}}\\
\multirow{4}{*}{$\mathrm{\alpha}$} & 0.9 &{\color{white}\graddb{0.76}}&{\color{white}\graddb{0.63}}& {\color{white}\graddb{0.61}}&{\color{white}\graddb{0.60}}\\
 & 0.63 &{\color{white}\graddb{0.69}}&{\color{white}\graddb{0.61}}&{\color{white}\graddb{0.60}}&{\color{white}\graddb{0.59}}\\
 & 0.37 &{\color{white}\graddb{0.64}}&{\color{white}\graddb{0.60}}&{\color{white}\graddb{0.58}}&{\color{white}\graddb{0.57}}\\
 & 0.1  &{\color{white}\graddb{0.63}}&{\color{white}\graddb{0.54}}&{\color{white}\graddb{0.53}}&{\color{white}\graddb{0.57}}\\
 & & 0.1 & 0.37 & 0.63 & 0.9\\
 & &\multicolumn{4}{c}{$\mathrm{\beta}$}\\
\end{tabular}
    \end{minipage} 

    \begin{minipage}{.5\linewidth}
\centering
\renewcommand{\arraystretch}{1.5}
\begin{tabular}[*{6}c]{*{6}c} 
 & &\multicolumn{4}{c}{\textbf{Dice-S{\o}rensen score - Fold 3}}\\
\multirow{4}{*}{$\mathrm{\alpha}$} & 0.9 &{\color{white}\graddc{0.94}}&{\color{white}\graddc{0.88}}& {\color{white}\graddc{0.87}}&{\color{white}\graddc{0.88}}\\
 & 0.63 &{\color{white}\graddc{0.92}}&{\color{white}\graddc{0.88}}&{\color{white}\graddc{0.87}}&{\color{white}\graddc{0.85}}\\
 & 0.37 &{\color{white}\graddc{0.70}}&{\color{white}\graddc{0.87}}&{\color{white}\graddc{0.85}}&{\color{white}\graddc{0.83}}\\
 & 0.1  &{\color{white}\graddc{0.87}}&{\color{white}\graddc{0.82}}&{\color{white}\graddc{0.80}}&{\color{white}\graddc{0.79}}\\
 & & 0.1 & 0.37 & 0.63 & 0.9\\
 & &\multicolumn{4}{c}{$\mathrm{\beta}$}\\
\end{tabular}
    \end{minipage} 
    \begin{minipage}{.5\linewidth}
\centering
\renewcommand{\arraystretch}{1.5}
\begin{tabular}[*{6}c]{*{6}c} 
 & &\multicolumn{4}{c}{\textbf{Dice-S{\o}rensen score - Fold 4}}\\
\multirow{4}{*}{$\mathrm{\alpha}$} & 0.9 &{\color{white}\graddd{0.62}}&{\color{white}\graddd{0.71}}& {\color{white}\graddd{0.68}}&{\color{white}\graddd{0.69}}\\
 & 0.63 &{\color{white}\graddd{0.66}}&{\color{white}\graddd{0.72}}&{\color{white}\graddd{0.74}}&{\color{white}\graddd{0.69}}\\
 & 0.37 &{\color{white}\graddd{0.70}}&{\color{white}\graddd{0.75}}&{\color{white}\graddd{0.69}}&{\color{white}\graddd{0.72}}\\
 & 0.1  &{\color{white}\graddd{0.74}}&{\color{white}\graddd{0.78}}&{\color{white}\graddd{0.78}}&{\color{white}\graddd{0.79}}\\
 & & 0.1 & 0.37 & 0.63 & 0.9\\
 & &\multicolumn{4}{c}{$\mathrm{\beta}$}\\
\end{tabular}
    \end{minipage} 
    
    \centerline{
    \begin{minipage}{.5\linewidth}
\centering
\renewcommand{\arraystretch}{1.5}
\begin{tabular}[*{6}c]{*{6}c} 
 & &\multicolumn{4}{c}{\textbf{Dice-S{\o}rensen score - Fold 5}}\\
\multirow{4}{*}{$\mathrm{\alpha}$} & 0.9 &{\color{white}\gradde{0.58}}&{\color{white}\gradde{0.67}}& {\color{white}\gradde{0.61}}&{\color{white}\gradde{0.61}}\\
 & 0.63 &{\color{white}\gradde{0.78}}&{\color{white}\gradde{0.62}}&{\color{white}\gradde{0.63}}&{\color{white}\gradde{0.58}}\\
 & 0.37 &{\color{white}\gradde{0.77}}&{\color{white}\gradde{0.61}}&{\color{white}\gradde{0.59}}&{\color{white}\gradde{0.57}}\\
 & 0.1  &{\color{white}\gradde{0.61}}&{\color{white}\gradde{0.56}}&{\color{white}\gradde{0.55}}&{\color{white}\gradde{0.55}}\\
 & & 0.1 & 0.37 & 0.63 & 0.9\\
 & &\multicolumn{4}{c}{$\mathrm{\beta}$}\\
\end{tabular}
    \end{minipage} 
    }
\caption{\rule{0pt}{3ex}Fold-wise Dice-Sørensen score for the networks evaluated on the related validation dataset, depending on the $\alpha$/$\beta$ parameters of the FTL.}
\label{tab:ps-folds}
\end{table}

\begin{table}
\centering
\renewcommand{\arraystretch}{1.4}
\begin{tabular}[*{11}c]{*{11}c} 
 & &\multicolumn{9}{c}{\textbf{Dice-S{\o}rensen score}}\\
\multirow{8}{*}{$\mathrm{\gamma}$}&2.0&{\color{white}\gradca{0.50}}& &{\color{white}\gradcb{0.60}}& &{\color{white}\gradcc{0.71}}& &{\color{white}\gradcd{0.52}}& &{\color{white}\gradce{0.64}}\\
&1.67&{\color{white}\gradca{0.45}}& &{\color{white}\gradcb{0.75}}& &{\color{white}\gradcc{0.82}}& &{\color{white}\gradcd{0.58}}& &{\color{white}\gradce{0.62}}\\
&1.5&{\color{white}\gradca{0.49}}& &{\color{white}\gradcb{0.75}}& &{\color{white}\gradcc{0.86}}& &{\color{white}\gradcd{0.61}}& &{\color{white}\gradce{0.80}}\\
&1.33&{\color{white}\gradca{0.67}}& &{\color{white}\gradcb{0.74}}& &{\color{white}\gradcc{0.84}}& &{\color{white}\gradcd{0.64}}& &{\color{white}\gradce{0.74}}\\
&1&{\color{white}\gradca{0.68}}& &{\color{white}\gradcb{0.81}}& &{\color{white}\gradcc{0.91}}& &{\color{white}\gradcd{0.80}}& &{\color{white}\gradce{0.87}}\\
&0.67&{\color{white}\gradca{0.72}}& &{\color{white}\gradcb{0.81}}& &{\color{white}\gradcc{0.90}}& &{\color{white}\gradcd{0.70}}& &{\color{white}\gradce{0.84}}\\
&0.5&{\color{white}\gradca{0.86}}& &{\color{white}\gradcb{0.69}}& &{\color{white}\gradcc{0.92}}& &{\color{white}\gradcd{0.66}}& &{\color{white}\gradce{0.78}}\\
&0.33&{\color{white}\gradca{0.75}}& &{\color{white}\gradcb{0.66}}& &{\color{white}\gradcc{0.90}}& &{\color{white}\gradcd{0.70}}& &{\color{white}\gradce{0.80}}\\
 & & 1 & & 2 & & 3 & & 4 & & 5\\
 & &\multicolumn{9}{c}{Fold}\\
\end{tabular}
\caption{\rule{0pt}{3ex}Fold-wise Dice-Sørensen score for the networks evaluated on the related validation dataset, depending on the $\gamma$ parameter of the FTL.}
\label{tab:ps-gamma-folds}
\end{table}
\end{appendices}

\end{document}